\documentclass[10pt, twocolumn]{article} 

\usepackage{paralist}
\usepackage{graphicx}
\usepackage{times}
\usepackage{latex8}
\setdefaultleftmargin{0.0em}{}{}{}{}{}
\renewenvironment{thebibliography}[1]{%
  \begin{oldthebibliography}{#1}%
    \setlength{\parskip}{0ex}%
    \setlength{\itemsep}{0ex}%
}
{
  \end{oldthebibliography}
}

\newcommand{\ProportionOfText}{0.47}

\begin{document}

\title{Reviewing Data Visualization: an Analytical Taxonomical Study}
\author{Jose F. Rodrigues Jr., Agma J. M. Traina, Maria Cristina F. de Oliveira, Caetano Traina Jr.\\
       Computer Science Department\\
       Computer and Mathematical Sciences Institute\\
       University of S\~ao Paulo at S\~ao Carlos\\
       Caixa Posta 668\\
       13560-970 S\~ao Carlos, SP, Brazil\\
       \{junio, agma, cristina, caetano\}@icmc.usp.br\\
       \\
       {\bf IEEE Copyright - http://ieeexplore.ieee.org/xpls/abs\_all.jsp?arnumber=1648338}
}

\maketitle
\thispagestyle{empty}

\subsection*{\centering Abstract}
{\em \noindent{ This paper presents an analytical taxonomy that can
suitably describe, rather than simply classify, techniques for data
presentation. Unlike previous works, we do not consider particular
aspects of visualization techniques, but their mechanisms and foundational vision perception. Instead of just adjusting
visualization research to a classification system, our aim is to
better understand its process. For doing so, we depart from
elementary concepts to reach a model that can describe how
visualization techniques work and how they convey meaning. }
}

\section{Introduction}
\label{Introduction}

\noindent{The large volume of data sets produced in all kinds of
human activities motivates the quest for more efficient ways to
explore and understand information. The benefits of such
understanding reflect in business advantages, more accurate
diagnosis, finer engineering and more refined conclusions in a
general sense. Computer graphics aided techniques have been
researched and implemented in order to provide improved mechanisms
for exploring stored data. These efforts are generically known as
(data) Visualization, which provides faster and user-friendlier
mechanisms for data analysis, because the user draws on his/her
comprehension immediately as graphical information comes up to
his/her vision.}

Several classification schemes have been proposed for visualization
techniques, each one focusing on some aspect of the visualization
process. However, many questions remain unanswered. What are the
building blocks of a visual exploration scene? How interaction
mechanisms relate to these facts? These are core issues for
implementing and evaluating visualization systems. In this work, we
discuss these issues and analytically find answers to them based on
the very mechanisms of the visualization techniques and on visual
perception theory.

In this paper we discuss the subjective nature of visualization by
proposing a discrete model that can better explain how visualization
scenes are composed and formed, and how their constituent parts
contribute to visual comprehension. We revisit visual analysis
proposing a perspective where visualization scenes are considered as
a set of components each of which passive of discrete consideration.
This discussion is organized as follows. Section \ref{RelatedWork}
reviews former taxonomies from the literature, section
\ref{Components} presents the basic components of visualization
techniques, used as elements for our proposed taxonomy. Section \ref{Our_Taxonomy_Proposal} delineates the
descriptive taxonomy itself, while section \ref{InteractionTechniques}
explains how interaction techniques fit into the proposed framework.
Finally, Section \ref{Conclusions} presents a brief discussion and
concludes the paper.

\section{Related Work}
\label{RelatedWork}

\noindent{One of the most referenced taxonomies for Visualization,
and well suited to academic purposes, is the one proposed by Keim
\cite{14}. It maps visualization techniques within a three
dimensional space defined by the following discrete axes: the data
type to be visualized (one, two, multi-dimensional, text/web,
hierarchies/graphs and algorithm/software), the visualization
technique (standard 2D/3D, geometrical, iconic, dense pixel and
stacked), and the interaction/distortion technique applied
(standard, projection, filtering, zoom, distortion and link \&
brush). This taxonomy is suitable to quickly reference and
categorize visualization techniques, but it is not adequate to
explain their mechanisms.}

A simpler taxonomy was earlier presented by Schneiderman \cite{29}.
It delineates a pair wise system based on a set of data types to be
explored, and on a set of exploratory tasks to be carried out by the
analyst. This taxonomy, known as task (overview, zoom, filter,
details-on-demand, relate, history and extract) by data type (one,
two, three, multi-dimensional, tree and network) taxonomy, was
pioneer in analytically delineating visualization techniques. The
effort provides a good idea of what a given technique is and how it
can be used.

Another interesting classification is presented by Chi \cite{5}, a
quite analytical approach, which details visualization techniques
through various properties related to a specific visualization
model. The taxonomy embraces data, abstraction, transformation and
mapping tasks, presentation and interaction. It determines a
complete and extensive descriptive system for analytical purposes.

Tory and M{\"o}oller \cite{32} define Scientific Visualization and
Information Visualization, respectively, as continuous ([one, two,
three, multi-dimensional] \textit{versus} [scalar, vector, tense,
multi-variate]) and discrete (two, three, multi-dimensional and
graph \& tree) classes, according to the intuitive perception of
their visual modeling. Wiss and Carr \cite{37} describe a cognitive
based taxonomy that considers attention, abstraction and
(interaction) affordance in order to discus 3-D techniques. Unlike a
classification system, this taxonomy can be seen as a guide to
``dissect" the subjective nature of visualization techniques.

In the following discussion we also analyze visualization techniques
according to a classification scheme but, differently, we
concentrate on basic characteristics common to every visually
informative scene. We benefit from empirical observations of how
data translates to space, that is, how it is \textit{spatialized}
and we consider characteristics proposed in visual perception theory
(position, shape and color), which determines how pre-attentive
features can stimulate our visual system.

\section{Components of Visualization Techniques}
\label{Components}

\begin{figure}[htb]
    \centering
\includegraphics[width=0.42\textwidth]{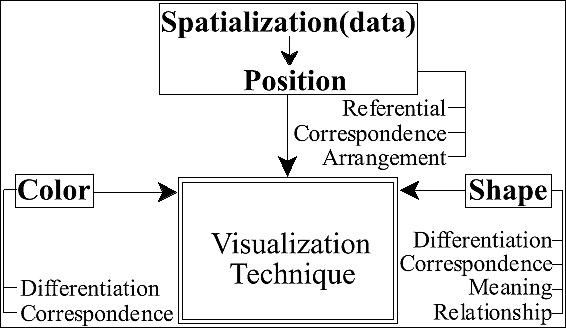}
    \caption{Taxonomy model based on spatialization and visual stimuli. The central rectangle represents visualization
    scenes. Around the visualization scene are its components: spatialization(position), shape and color together with their possible classes.
    }
    \label{Taxonomy_Model}
\end{figure}

\noindent{Visualization can be understood as data represented
visually. That is, it takes advantage of {\it spatialization} to
allow data to be visually/spatially perceived and it relies on {\it
visual stimuli} to represent data itens or data
attributes/characteristics. Based on these facts an overview of our
taxonomy model is presented in Figure \ref{Taxonomy_Model}, which
depicts its basic components and their possible classes, further
detailed in this text.}

\subsection{Spatialization}

\noindent{Spatialization of data refers to its transformation from a
raw format that is difficult to interpret into a visible spatial
format. In fact, Rohrer {\it et al.} \cite{27} state that
visualizing the non-visual requires mapping the abstract into a
physical form, and Rhyne {\it et al.} \cite{25} differentiate
Scientific visualization and Information visualization based on whether the spatialization mechanism is given or chosen, respectively. We considered these arguments to analyze spatialization and verified
that visualization techniques can be grouped based on how they
are mapped into the visual/spatial domain.}

\subsection{Pre-attentive Visual Stimuli}

\noindent{Semiotic theory is the study of signs and how they convey
meaning. According to semiotic theory, the visual process is
comprised of two phases, the parallel extraction of low-level
properties (called pre-attentive processing) followed by a
sequential goal-oriented slower phase. Pre-attentive processing
plays a crucial role in promoting visualization's major gain, that
is, improved and faster data comprehension \cite{43}.}

\begin{figure}[htb]
    \centering
\includegraphics[width=\ProportionOfText\textwidth]{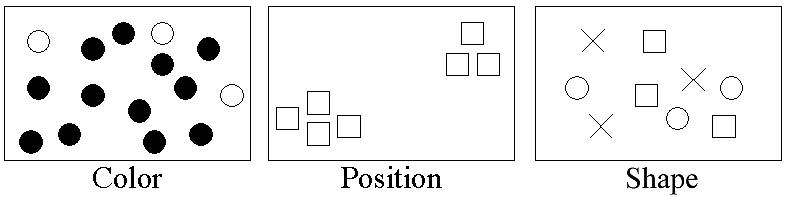}
    \caption{Pre-attentive visual stimuli.}
    \label{Semiotical}
\end{figure}

Specifically, pre-attentive processing refers to what can be
visually identified prior to conscious attention. Essentially, it
determines which visual objects are instantly and effortlessly
brought to our attention. The work described by Ware \cite{35} identifies the categories of visual features that are pre-attentively processed. Position (2D position, stereoscopic depth, convex/concave shading), Shape (line orientation, length, width and line collinearity, size, curvature, spatial grouping, added marks, numerosity) and Color (hue, saturation) are considered and, according to Pylyshyn {\it et al} \cite{23}, specialized areas of the brain exist to process each of them (Figure \ref{Semiotical}). Actually this is true for everything we see, for what we can ask three questions: where is it? what is its shape? and what color is it?

\section{Proposed Taxonomy}
\label{Our_Taxonomy_Proposal}

\noindent{Visualizing data demands a maximization of just noticeable differences. To satisfy this need, visualizations rely on pre-attentive stimuli - characteristics inherent to visual/spatial entities. Therefore, the data must
first be mapped to the spatial domain (spatialized) in order to be
pre-attentively perceived. Our
taxonomy thus focuses on the spatialization process and on the
pre-attentive stimuli that are employed by visualization techniques.}

\subsection{Spatialization}

\noindent{In this section we identify a set of procedures for data
spatialization: Structure exposition, Projection, Patterned
positioning and Reproduction. In the following section we present
the pre-attentive stimuli that complete the requisites to describe a
data visualization.}

\begin{compactitem}

    \item{{\it Structure exposition}: data can embed intrinsic structures,
    such as hierarchies or relationship networks (graph-like),
    that embody a considerable part of the data significance. This class comprises visualization
    techniques that rely on methods to adjust data presentation so that the underlying
    data structure can be visually perceived. Examples are the TreeMap technique \cite{30}, illustrated
    in Figure \ref{StructureExposition}(a), and force directed graph layouts \cite{41}, such as the one illustrated in
    Figure \ref{StructureExposition}(b)};

\begin{figure}[htb]
    \centering
\includegraphics[width=\ProportionOfText\textwidth]{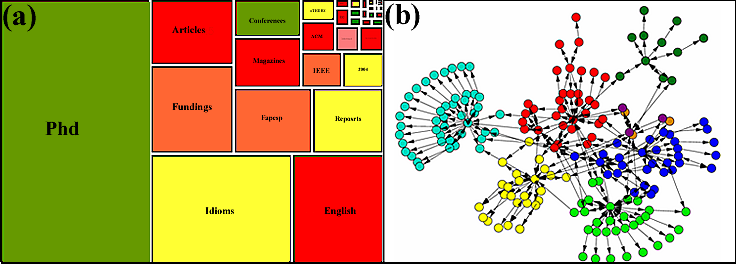}
    \caption{(a) TreeMap structure exposition. {\it Position:} hierarchical arrangement;
    {\it shape:} correspondence (size proportionality); {\it color:} discrete differentiation.
    (b) Force directed structure exposition. {\it Position:} relational arrangement;
    {\it shape:} meaningful (arrowed) lines; {\it color:} discrete differentiation.}
    \label{StructureExposition}
\end{figure}

    \item{{\it Patterned}: this is the simplest positioning procedure, with the set of individual data items arranged sequentially (ordered or not) according to one or more directions, linear, circular or according to specific patterns.
  }

Patterned techniques tend to fully populate the projection area and
sometimes are referred to as dense pixel displays. Examples include
Pixel Bar Charts \cite{16}, showed in Figure \ref{Patterned}(a), pie
charts (circular disposition), depicted in \ref{Patterned}(b) and
pixel oriented techniques in general \cite{13}.

\begin{figure}[htb]
    \centering
\includegraphics[width=\ProportionOfText\textwidth]{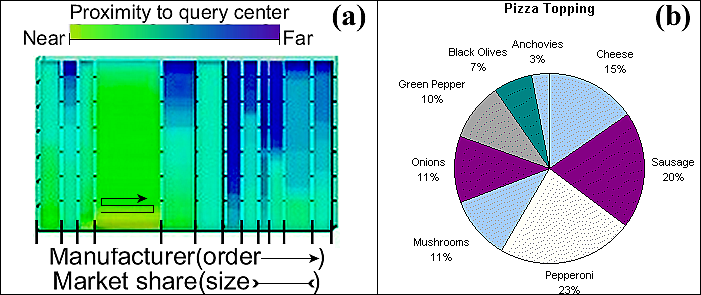}
    \caption{(a) Pixel Bar Charts. {\it Position:} manufacturers mapped in horizontal sequence
    and clients (pixels) mapped according to a patterned positioning; {\it shape:} correspondence
    (size); {\it color:} continuous correspondence.
    (b) Pie chart. {\it Position:} each slice maps a different pizza ingredient;
    {\it shape:} correspondence (size); {\it color:} discrete differentiation.}
    \label{Patterned}
\end{figure}

Notice that the simple approach of Patterned positioning restricts
the presentation of data, which is tipically depicted with shape and
size encoding, as in Figures \ref{Patterned}(a), \ref{Patterned}(b)
and \ref{ShapePosition}(c). Keim's pixel-oriented techniques are an
exception, in that they use just color, and no shape encoding, to
present the data items, which are positioned according to elaborated
patterned sequences \cite{15}.

\begin{figure}[htb]
    \centering
\includegraphics[width=\ProportionOfText\textwidth]{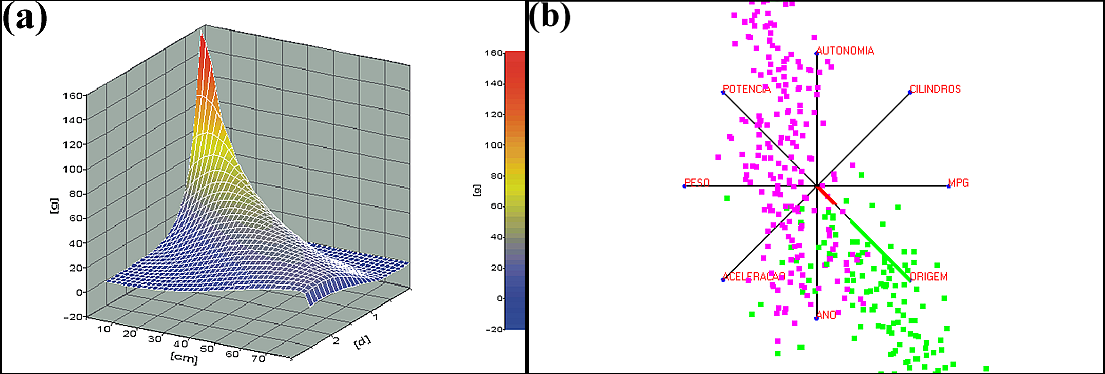}
    \caption{3D functional plotting. {\it Position:} referential (axes); {\it shape:} math surface displays
    relationship among points; meaningful axes and labels; {\it color:} continuous correspondence. (b) Star Coordinates
    2D projection of 8-dimensional data. {\it Position:} referential; {\it shape:} meaningful axes and labels; {\it color:}
discrete differentiation.}
    \label{Projection}
\end{figure}

    \item{{\it Projection}: stands for a data display modeled by the representation
    of functional variables. That is, the position of a data item is defined by
    either a well-known or an implicit mathematical function. In a projection, the information
    given is magnitude and not order, as in a patterned spatialization. Examples are
    Parallel Coordinates (one projection per data dimension), Star Coordinates \cite{12} and conventional
    graph plots, as illustrated in Figure \ref{Projection};

    }

    \item{{\it Reproduction}: data positioning is known beforehand and is determined by the
    spatialization of the system from where data were collected, as exemplified in Figures \ref{Reproduction}(a)
    and \ref{Reproduction}(b). In reproduction, the data inherits positioning from its
    original source. Usually, specific algorithms \cite{40} are required to identify the data positioning based on
    its implicit physical structure; other algorithms are used to simplify intractable volumes and/or to derive
    other features later represented for example as color, glyphs or streamlines.

\begin{figure}[htb]
    \centering
\includegraphics[width=\ProportionOfText\textwidth]{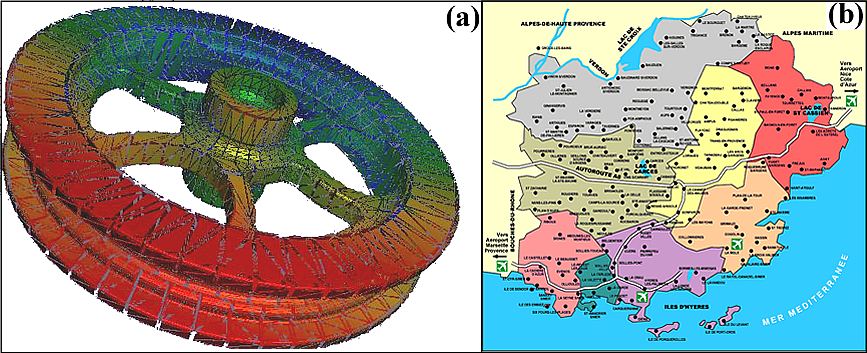}
    \caption{(a) Rendered dataset. {\it Position:} referential (surface shape as reference); {\it shape:}
    meaningful object; {\it color:} continuous correspondence.
    (b) Geographical map. {\it Position:} referential
    (background map as reference); {\it shape:} meaningful airport and road identifiers; {\it color} discrete
    differentiation. \small{(b) reproduced with permission granted by S.G. Eick.}}
    \label{Reproduction}
\end{figure}
    }
\end{compactitem}

Reproduction can be seen as a special case of projection where no
explicit projection function is given - compare Figures
\ref{Projection}(a) and \ref{Reproduction}(a). Instead, the data
positioning derives from the observed phenomenon and it is part of
the data. Therefore, projection and reproduction are characterized
by considerably different methodologies that confer them distinct
classifications, namely, explicit projection and implicit projection
(rendering).

\subsection{Pre-attentive Stimuli}

\noindent{In this section we analyze well-known visualization
techniques in order to empirically identify how attributes Position,
Shape and Color are used to express information.}

\subsubsection{Position}

\noindent{Position is the primary component for pre-attention
perception in visualization scenes and it is strictly related to the
spatialization process. So, while spatialization is the cornerstone
for enabling visual data analysis (as it maps data to the
visual/spatial domain), it also dictates the mechanism for
pre-attentive positional perception. Thus, positional pre-attention
occurs in the form of Arrangement, Correspondence and Referential,
explained in the following paragraphs. These classes derive,
respectively, from spatializations Structure Exposition, Patterned
and Projection/Reproduction.}

\begin{figure}[htb]
    \centering
\includegraphics[width=\ProportionOfText\textwidth]{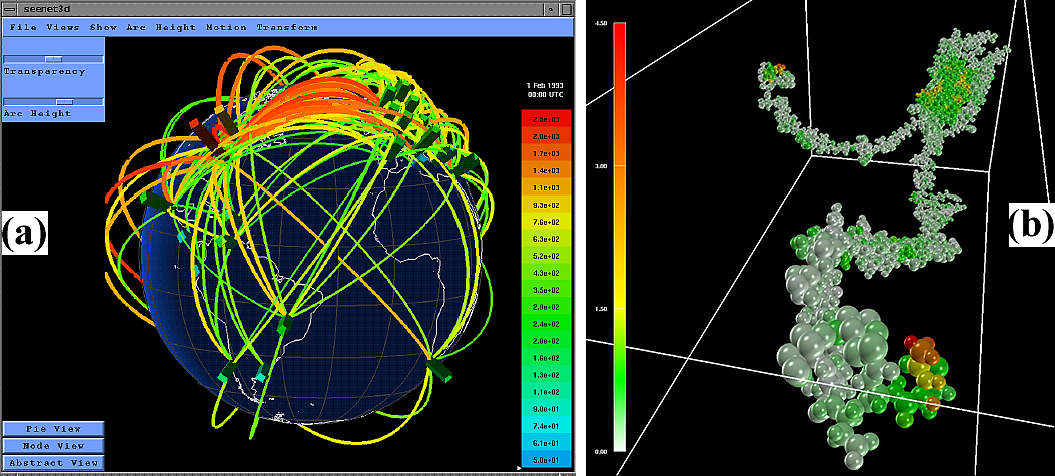}
    \caption{(a) {\it Position}: referential (globe as reference); {\it shape}: relationship (curved lines), proportional
    correspondence (pillars size) and meaning (globe); {\it color}: discrete correspondence. (b) {\it Position}:
    referential (parallelepiped as reference); {\it shape}: meaning (chemical molecules); {\it color}: continuous
    correspondence. \small{Images reproduced with permission granted by S.G. Eick.}}
    \label{Position_ab}
\end{figure}

\begin{compactitem}
    \item{{\it Structure Exposition $\rightarrow$ Arrangement}: specific arrangements can depict
    structure, hierarchy or some other global property. Without an explicit referential, information
    is perceived locally through individual inter-positioning of elements and/or globally through
    a scene overview. For instance, TreeMap (Figure \ref{StructureExposition}(a)) presents
    the hierarchy of the data items, and a graph layout (Figure \ref{StructureExposition}(b)) presents
    network information.}

    \item{{\it Patterned $\rightarrow$ Correspondence}: the position of an item, either discrete or continuous,
    determines its corresponding characteristic without demanding a reference. For example,
    see Figure \ref{ShapePosition}(b) where each of the four line positions maps one data attribute.
    Other examples are Parallel Coordinates and Table Lens \cite{24}, techniques that define
    an horizontal sequence for placing data attributes;}

    \item{{\it Projection $\rightarrow$ Referential}: this is the most obvious relation between spatialization
    and positional pre-attention. Projections have a supporting function whose intervals define referential scales
    suited to analogical comprehension.}

    \item{{\it Reproduction $\rightarrow$ Referential}: the position of an element, discrete or continuous, is
    given relative to an explicit reference, such as a geographical map (Figure \ref{Reproduction}(b)), a
    meaningful shape (Figure \ref{Position_ab}(a)) or a set of axes (Figure \ref{Position_ab}(b));}

\end{compactitem}

We observe that spatialization based on {\it Reproduction} can yield
all the three kinds of positional pre-attention: arrangement,
correspondence or referential. This is a consequence of the image
characteristics being predetermined from the source being
reproduced, which may bear any of these characteristics naturally.
However, the most common occurrence is referential pre-attention, to
which we limit our exposition.

\subsubsection{Shape}

\noindent{We argued so far that a limited number of spatialization
procedures is at the core of visualization techniques, and that
these spatialization procedures dictate the positional pre-attentive
stimulus. Nevertheless, after spatializing the data one still needs
to choose their shape and color, other pre-attentive stimuli. In
this and the following sections we investigate how shape and color
contribute to visual perception. In particular, the Shape stimulus
embraces the largest number of possibilities to express information:
Differentiation, Correspondence, Meaning and/or Relationship.}

\begin{figure}[htb]
    \centering
\includegraphics[width=\ProportionOfText\textwidth]{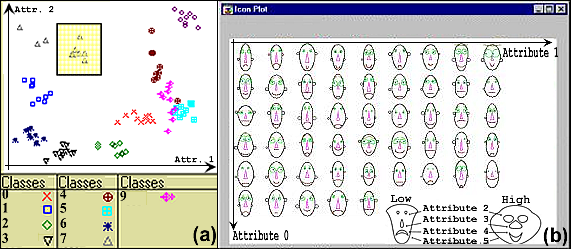}
    \caption{(a) {\it Position}: referential (axes as reference); {\it shape}: discrete correspondence and differentiation (square); {\it color}: discrete correspondence. (b) {\it Position}: referential (axes as reference); {\it shape}: continuous correspondence (size and curvature); {\it color}: discrete differentiation.}
    \label{Shape_ab}
\end{figure}

\begin{compactitem}

    \item{{\it Differentiation}: the shape displayed discriminates the items for further interpretation,
    as in Figures \ref{Shape_ab}(a), \ref{Color_ab}(a) and \ref{ShapePosition}(a);}
    \item{{\it Correspondence}: discrete (Figure \ref{Shape_ab}(a)) or continuous (Figure \ref{Shape_ab}(b)),
    each noticeable shape corresponds to one informative feature. Proportion (variable sizing) is the most
    used variation for this practice;}
    \item{{\it Meaning}: the shape displayed carries meaning, such as an arrow, a face or a complex shape (e.g. text),
    whose comprehension may depend on user's knowledge and previous experience, as depicted in Figures
    \ref{Position_ab}(b) and \ref{Shape_ab}(b);}
    \item{{\it Relationship}: shapes, such as lines, contours or surfaces, denote the relationship between
    a set of data items, e.g., in Parallel Coordinates, 3D plots and paths in general, illustrated in
    Figures \ref{Projection}(a) and \ref{Position_ab}(a).}

\end{compactitem}

\subsubsection{Color}

\noindent{After applying a spatialization procedure, which leads to
positional clues for perceiving information, and after choosing a shape to
convey additional meaning, color is the third pre-attentive stimulus to be considered. Color conveys information by Differentiation and/or
Correspondence of data items:}

\begin{figure}[htb]
    \centering
\includegraphics[width=\ProportionOfText\textwidth]{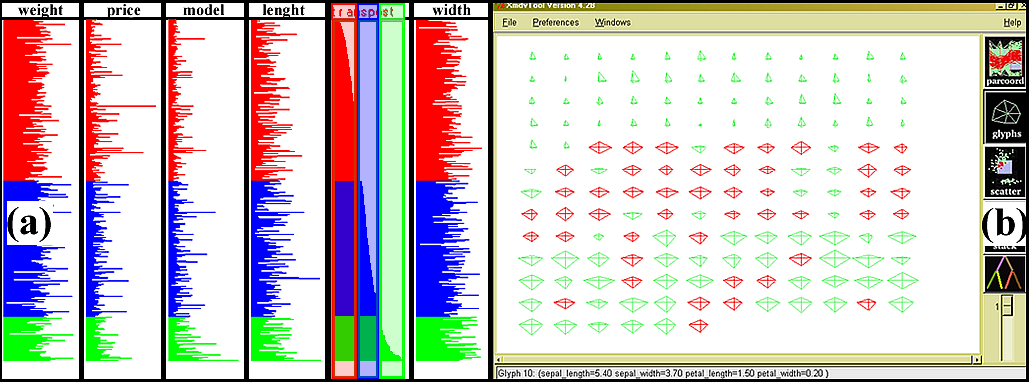}
    \caption{(a) {\it Position}: correspondence for attribute order and attribute values; {\it shape}: line size
    correspondence (all columns) and differentiation (in the 5th column shape indicates selection); {\it color}: discrete
differentiation. (b) {\it Position}:
    referential (window as a Euclidean plane) and correspondence through the circular positioning of the inner
    sticks of each glyph; {\it shape}: differentiation determined by the contour around each glyph and proportional correspondence
    for the inner sticks; {\it color}: discrete differentiation. \small{(b) created with XmdvTool \cite{42}}.}
    \label{Color_ab}
\end{figure}

\begin{compactitem}

    \item{{\it Differentiation}: colors have no specific data correspondence, they just depict equality
    (or inequality) of some data characteristic, as it may
    be observed in the visualizations depicted in Figures \ref{Color_ab}(a) and \ref{Color_ab}(b). The coloring of the items,
    either discrete or continuous, is data dependent or user input dependent;}
    \item{{\it Correspondence}: discrete or continuous, as observed in Figures \ref{Position_ab}(a) and (b). In
    the discrete case each noticeable color maps one informative feature, usually a class, a level,
    a stratum or some predefined correspondence. In the continuous case, the variation of tones maps
    a set of continuous data values.}

\end{compactitem}

\subsection{Hybridism and Subspace Visualizations}
\label{Hybridism}

\begin{figure*}[htb]
    \centering
\includegraphics[width=0.75\textwidth]{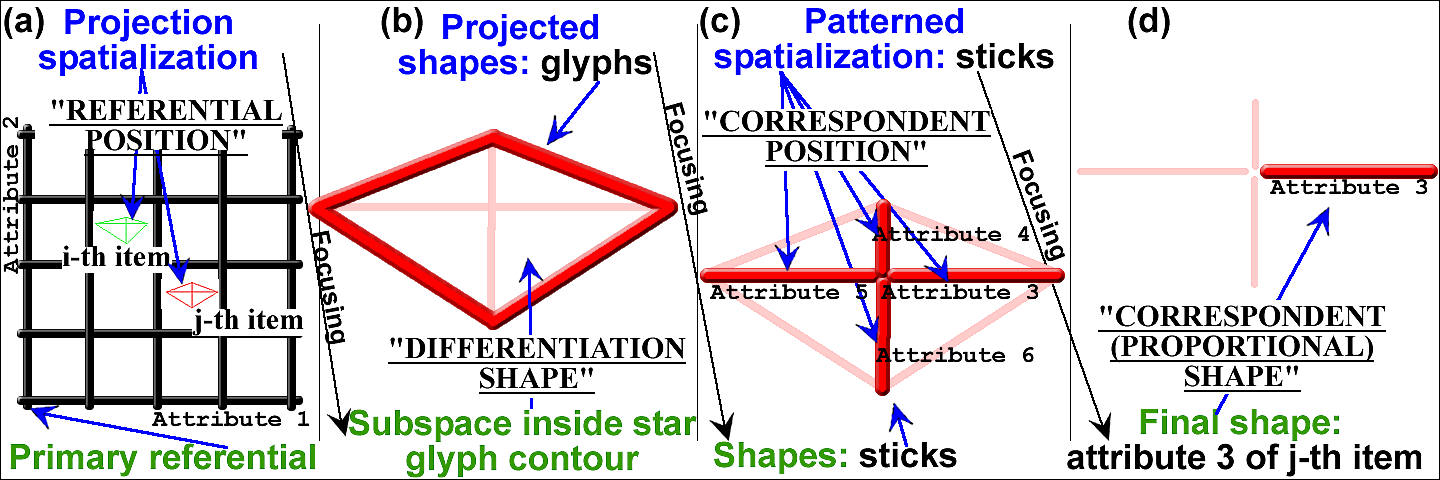}
    \caption{Two spatialization cycles applied in generating a visualization.
    (a) Projection of star glyphs. (b) Star glyph focused. (c) Arrangement within a glyph.
    (d) Attribute information as a proportional shape.}
    \label{ShapePosition}
\end{figure*}

\noindent{In the process of creating a visualization, it is possible
to subdivide the available space into disjoint regions and, then,
apply another spatialization process to each subspace. Figure
\ref{ShapePosition}(a), for example, shows a grid in which star
glyphs are spatialized. Similarly, Figure \ref{ShapePosition}(b)
shows a focused star glyph in which sticks are positioned according
to a different spatialization procedure. Figure
\ref{ShapePosition}(c) demonstrates the relative positioning of the
glyphs in the star glyph and, finally, Figure \ref{ShapePosition}(d)
shows a focused stick that represents the magnitude of the third
attribute of the (hypothetical) jth item.}

A similar approach is applied in techniques such as Dimensional
Stacking \cite{18}, Worlds-within-Worlds \cite{7}, Circle Segments
\cite{1}, Pixel Bar Charts \cite{16} and the so-called iconic
techniques in general. Multiple spatialization cycles allow improved
space utilization and result in more complex visualization
techniques. Moreover, they define hybrid approaches for composing
visualizations that allow for the vast number of techniques found in
visualization literature. In such compositions, pre-attention occurs
as a function of the visualization focus. Such understanding,
coupled with our taxonomical system, can provide additional guidance
on new thoughts for data visualization.

\hyphenpenalty=10000
\begin{table*}[th]
  \begin{center}
  \caption{Examples of spatialization-pre-attention analysis}
  \label{ClassificationExamplesTable}
  \begin{tabular}{|p{1.3in}|p{1.7in}|p{0.7in}|p{0.7in}|p{1.1in}|}
    \hline    	 
          {\footnotesize{\it{Visualization Technique}}}
         &{\footnotesize{\it{Spatialization$\rightarrow$Position}}}
         &{\footnotesize{\it{Shape}}}
         &{\footnotesize{\it{Color (usual)}}}
         &{\footnotesize{\it{Prospective Interaction}}}\\
    \hline
    \hline    
      {\footnotesize{Chernoff Faces \cite{4}}}
     &{\footnotesize{Projection$\rightarrow$Referential, Patterned$\rightarrow$Correspondence}}
     &{\footnotesize{Correspondence, Differentiation}}
     &{\footnotesize{-}}
     &{\footnotesize{Filtering}}\\
    \hline 
      {\footnotesize{Dimensional Stacking \cite{18}}}
     &{\footnotesize{Projection$\rightarrow$Referential}}
     &{\footnotesize{-}}
     &{\footnotesize{Differentiation}}
     &{\footnotesize{Filtering}}\\
    \hline 
      {\footnotesize{Parallel Coordinates \cite{10}}}
     &{\footnotesize{Projection$\rightarrow$Referential, Patterned$\rightarrow$Correspondence}}
     &{\footnotesize{Relationship}}
     &{\footnotesize{Differentiation}}
     &{\footnotesize{Filtering}}\\
    \hline
      {\footnotesize{Scatter Plots \cite{6}}}
     &{\footnotesize{Projection$\rightarrow$Referential}}
     &{\footnotesize{-}}
     &{\footnotesize{Differentiation}}
     &{\footnotesize{Filtering}}\\
    \hline
      {\footnotesize{Star Coordinates \cite{12}}}
     &{\footnotesize{Projection$\rightarrow$Referential, Patterned$\rightarrow$Correspondence}}
     &{\footnotesize{-}}
     &{\footnotesize{Differentiation}}
     &{\footnotesize{\mbox{View transformation}, \mbox{Details-on-demand}}}\\
    \hline
       {\footnotesize{Stick Figures \cite{22}}}
      &{\footnotesize{Projection$\rightarrow$Referential, Patterned$\rightarrow$Correspondence}}
      &{\footnotesize{Differentiation, Correspondence}}
      &{\footnotesize{Differentiation}}
      &{\footnotesize{Filtering}}\\
    \hline
       {\footnotesize{Worlds-within-Worlds \cite{7}}}
      &{\footnotesize{Projection$\rightarrow$Referential}}
      &{\footnotesize{-}}
      &{\footnotesize{-}}
      &{\footnotesize{\mbox{View transformation}}}\\
    \hline
      {\footnotesize{Bar Chart}}
     &{\footnotesize{Projection$\rightarrow$Referential}}
     &{\footnotesize{Correspondence}}
     &{\footnotesize{Correspondence}}
     &{\footnotesize{Filtering, Parametric}}\\
    \hline
      {\footnotesize{Pixel Bar Charts \cite{16}}}
     &{\footnotesize{Projection$\rightarrow$Referential, Patterned$\rightarrow$Correspondence}}
     &{\footnotesize{Correspondence}}
     &{\footnotesize{Correspondence}}
     &{\footnotesize{Filtering, Parametric}}\\
    \hline
      {\footnotesize{Circle Segments \cite{1}}}
     &{\footnotesize{Patterned$\rightarrow$Correspondence}}
     &{\footnotesize{-}}
     &{\footnotesize{Correspondence}}
     &{\footnotesize{\mbox{Details-on-demand}}}\\
    \hline
      {\footnotesize{Keim's Pixel Oriented \cite{13}}}
     &{\footnotesize{Patterned$\rightarrow$Correspondence}}
     &{\footnotesize{-}}
     &{\footnotesize{Differentiation}}
     &{\footnotesize{Filtering, Parametric}}\\
    \hline
      {\footnotesize{Pie Chart}}
		 &{\footnotesize{Patterned$\rightarrow$Correspondence}}
		 &{\footnotesize{Correspondence}}
		 &{\footnotesize{Correspondence}}
		 &{\footnotesize{Filtering, Parametric}}\\
    \hline
      {\footnotesize{Table Lens \cite{24}}}
     &{\footnotesize{Patterned$\rightarrow$Correspondence, Projection$\rightarrow$Referential}}
     &{\footnotesize{Correspondence}}
     &{\footnotesize{Differentiation}}
     &{\footnotesize{Filtering, \mbox{Details-on-demand}}}\\
    \hline
      {\footnotesize{Cone Tree \cite{26}}}
     &{\footnotesize{Structure Exposition$\rightarrow$Arrangement}}
     &{\footnotesize{Relationship}}
     &{\footnotesize{Differentiation}}
     &{\footnotesize{\mbox{View transformation}, \mbox{Details-on-demand}}}\\
    \hline
      {\footnotesize{Hyperbolic Tree \cite{17}}}
     &{\footnotesize{Structure Exposition$\rightarrow$Arrangement}}
     &{\footnotesize{Relationship}}
     &{\footnotesize{Differentiation}}
     &{\footnotesize{\mbox{View transformation}, \mbox{Details-on-demand}}}\\
    \hline
      {\footnotesize{Treemaps \cite{30}}}
     &{\footnotesize{Structure Exposition$\rightarrow$Arrangement}}
     &{\footnotesize{Correspondence}}
     &{\footnotesize{Differentiation}}
     &{\footnotesize{Filtering, \mbox{Details-on-demand}}}\\
    \hline    
      {\footnotesize{Geographical Maps}}
     &{\footnotesize{Reproduction$\rightarrow$Referential}}
     &{\footnotesize{Differentiation, Correspondence}}
     &{\footnotesize{Differentiation}}
     &{\footnotesize{Filtering, \mbox{Details-on-demand}}}\\
    \hline
      {\footnotesize{Vector Visualization}}
     &{\footnotesize{Reproduction$\rightarrow$Referential}}
     &{\footnotesize{Meaning, Correspondence}}
     &{\footnotesize{Differentiation, Correspondence}}
     &{\footnotesize{\mbox{View transformation}}}\\
    \hline
      {\footnotesize{\mbox{Direct Volume Rendering} \cite{36}}}
     &{\footnotesize{Reproduction$\rightarrow$Referential}}
     &{\footnotesize{Meaning}}
     &{\footnotesize{Differentiation, Correspondence}}
     &{\footnotesize{\mbox{View transformation}}}\\
    \hline    
	\end{tabular}
  \end{center}	
\end{table*}	
 \hyphenpenalty=50

\section{Interaction Techniques}
\label{InteractionTechniques}

\noindent{Interaction is an important component for visualization
techniques but, differently from former works, we do not incorporate
interaction to our taxonomy. In fact, we chose to handle
visualization and interaction as disjoint concepts. However,
interaction and visual applications present a notable synergy.
Therefore, we must clarify the role of interaction techniques in the
visualization scene. We define two conditions for identifying an
interaction technique:}

\begin{enumerate}

    \item{An interaction technique must enable a user to define/redefine
    the visualization by modifying the characteristics of pre-attentive stimuli;}

    \item{An interaction technique, with appropriate adaptations, must be applicable to
    any visualization technique.}

\end{enumerate}

The first condition arises from the direct assumption that
interaction techniques alter the state of a computational
application. In the case of a visualization scene, its basic
components (the pre-attentive stimuli) must be altered. The second
condition arises from the need of having interaction techniques that
are well defined, which directs us towards generality. An
interaction technique, then, must be applicable to any visualization
technique, even if not efficiently. We identify the following
interaction paradigms:

\begin{compactitem}

    \item{{\it Parametric}: the visualization is redefined, visually (e.g., scrollbar) or textually (e.g., type-in),
    by modifying position, shape or color parameters. One could mention as examples, the hierarchical Parallel
    Coordinates (visual) mechanism described by Fua {\it et al} \cite{8} and Keim's \cite{13} query-dependent pixel
    displays that conform to a textual query system;}
    \item{{\it View transformation}: this interaction adds physical touch to the visualization scene, whose shape (size)
    and position can be changed through scale, rotation, translation and/or zoom, not necessarily all of them, as
    in the FastmapDB tool \cite{33};}
    \item{{\it Filtering}: a user is allowed to visually select a subset of items that, through pre-attentive factors such as
    color (brushing) and shape (selection contour), will be promptly differentiated for user perception. Detailed studies
    are presented by Martin and Ward \cite{20};}
    \item{{\it Details-on-demand}: detailed information about the data that generated a particular visual entity can be retrieved at
    any moment and presented in the scene. As an example we refer to the interaction (not the presentation) of
    Table Lens visualization, which permits to retrieve the data that originated a given graphical item and present it in textual (shape) form;}
    \item{{\it Distortion}: allows visualizations to be projected so that different perspectives (positions) can be
    observed and defined simultaneously. Classical examples are the Perspective Wall \cite{19} and Fish-eye
    Views \cite{28}.}

\end{compactitem}

The well-known Link \& Brush (co-plots) technique does not satisfy
our conditions. It is an application dependent automation that can
be implemented only when brushing and multiple visualization
techniques share a visualization environment.

\vspace{-0.4cm}
\section{Conclusions}
\label{Conclusions}
\vspace{-0.4cm}
\noindent{We illustrate the proposed taxonomy in Table
\ref{ClassificationExamplesTable}, which shows the categorization of
some well-known visualization techniques. In proposing this taxonomy
we focused on generalizing the rationale of how visualization scenes
are engendered and how they are presented to our cognitive system.
Such a general characterization results in a taxonomy that does not
rely on specific details on how techniques operate. Rather, it
considers their fundamental constituent parts: how they perform
spatialization and how they employ pre-attentive stimuli to convey meaning.
Our claim is that such an approach is required to gain a general
understanding of the visualization process.}

Existing taxonomies categorize techniques based on diverse and
detailed information on how techniques perform a visual mapping.
This diversity and detailing (refer to Section \ref{RelatedWork})
include, e.g., axes arrangement (``stacked techniques"), specific
representational patterns (``iconic and pixel-oriented techniques"),
predisposition of representativeness (``network and tree
techniques"), dimensionality (``2D/3D techniques") and interaction
(``static/dynamic techniques"). Although such approaches can
suitably describe the set of available techniques, they lack
analytical power because the core constituents of the techniques are
diffused within the taxonomical structure.

Our approach results in an extensible taxonomy that can accommodate
new techniques as, in fact, any technique will rely on common
foundational basis. We see this taxonomy as a starting point for
fomenting further discussions and thoughts on how visualization
techniques operate and how we can improve our understanding of them.
Hopefully, it can contribute to give us better
grounds for design, evaluation and implementation of techniques in the future. \\

\vspace{-0.4cm}
\noindent{\textbf{Aknowledgements} This work has been supported by
FAPESP (S\~ao Paulo State Research Foundation), CNPq (Brazilian
National Research Foundation) and CAPES (Brazilian Committee for
Graduate Studies).

\bibliographystyle{plain}
  {\small
    {

    }
}

\end{document}